\documentclass[10pt]{article}

\usepackage{times}
\usepackage{graphicx}
\usepackage{amsmath,amssymb}
\usepackage{multirow}
\usepackage{caption}
\usepackage{threeparttable}
\usepackage [latin1]{inputenc}

\newenvironment{sciabstract}{%
\begin{quote} }
{\end{quote}}

\newcounter{lastnote}

\title{An Agile Very Low Frequency Radio Spectrum Explorer}

\author
{Linjie~Chen,$^{1\ast}$ Yihua~Yan,$^{1,2}$ Qiuxiang~Fan,$^{3}$ Lihong Geng,$^{1}$ S. K. Bisoi,$^{1,4}$\\
\\
\footnotesize{$^{1}$Key Laboratory of Solar Activity, National Astronomical Observatories, Chinese Academy of Sciences, Beijing 100012, China}\\
\footnotesize{$^{2}$School of Astronomy and Space Sciences, University of CAS, Beijing 100049, China}\\
\footnotesize{$^{3}$Institute of Automation, Chinese Academy of Sciences, 100190, Beijing, China}\\
\footnotesize{$^{4}$National Institute of Technology, Rourkela, 769008, India}\\
\footnotesize{$^\ast$Corresponding author; E-mail:  ljchen@nao.cas.cn.}
}

\date{Oct. 10, 2020}

\begin{document}


\baselineskip18pt


\maketitle


\begin{sciabstract}
The very low frequency (VLF) regime below 30 MHz in the electromagnetic spectrum has presently drawing global attentions in radio astronomical research due to its potentially significant science outcomes exploring many unknown extragalactic sources, transients, and so on. However, the non-transparency of the Earth's ionosphere, ionospheric distortion and artificial radio frequency interference (RFI) have made it difficult to detect the VLF celestial radio emission with ground-based instruments. A straightforward solution to overcome these problems is a space based VLF radio telescope, just like the VLF radio instruments onboard the Chang'E 4 spacecraft. But building such a space telescope would be inevitably costly and technically challenging. The alternative approach would be then a ground based VLF radio telescope. Particularly, in the period of post 2020 when the solar and terrestrial ionospheric activities are expected to be in a 'calm' state, it will provide us a good chance to perform VLF ground-based radio observations. Anticipating such an opportunity, we built an agile VLF radio spectrum explorer co-located with the currently operational Mingantu Spectra Radio Heliograph (MUSER). The instrument includes four antennas operating in the VLF frequency range 1--70 MHz. Along with them, we employ an eight-channel analog and digital receivers to amplify, digitize and process the radio signals received by the antennas. We present in the paper this VLF radio spectrum explorer and the instrument will be useful for celestial studies of VLF radio emissions.
\end{sciabstract}

% In setting up this template for *Science* papers, we've used both
% the \section* command and the \paragraph* command for topical
% divisions.  Which you use will of course depend on the type of paper
% you're writing.  Review Articles tend to have displayed headings, for
% which \section* is more appropriate; Research Articles, when they have
% formal topical divisions at all, tend to signal them with bold text
% that runs into the paragraph, for which \paragraph* is the right
% choice.  Either way, use the asterisk (*) modifier, as shown, to

\section{Introduction}           %% first-level sections will be auto-capitalized
\label{sect:intro}
The VLF regime is the last possible unexplored regime to be studied in the whole electromagnetic (EM) spectrum in order to reveal several important astronomical sciences such as dark matter, galactic interstellar medium, transients and variable sources, etc (\cite{heino2009},~\cite{boonstra2016}). However, due to limitation of the Earth's ionosphere and strong artificial radio frequency interference (RFI), the celestial radio emissions below 30 MHz are not studied in details with ground based radio telescopes. To study the low frequency universe several ground based radio telescopes have been built to date, such as the Low Frequency Array (LOFAR) in Europe \cite{haarlem2013}, the Long Wavelength Array (LWA) in New Mexico, USA \cite{ellingson2009}, and the Giant Ukrainian Radio Telescope in Ukraine \cite{konovalenko2016}. However, they mostly operate above 10 MHz and are severely limited in radio observations between 10 to 30 MHz.

In order to explore the VLF spectral regime, building radio telescopes locating beyond the Earth's ionosphere is the most viable solution. But the telescopes should be either located far way from the Earth or the observations from them should be shielded from the Earth-based RFI so that the RFI impact from the Earth can be reduced to the minimum level (\cite{alexander1975},~\cite{kaiser1996}). Keeping these in mind, several space-based instruments have already been built to carry out the VLF observations (\cite{herman1973},~\cite{kaiser1996},~\cite{gopalswamy2014}) such as the very low frequency radio spectrometer (\cite{ji2017}) and the Netherlands-China Low Frequency Explorer (NCLE) onboard Chang'E-4 mission (\cite{boonstra2017}), which is the latest radio instrument operating in this field. Nevertheless, all of these instruments are mostly single element types, and it is also costly and technologically challenging to develop such space-based instruments.

The solar activity varies periodically in every 11 years and it is known as solar cycle. Currently, we are experiencing the minimum of solar cycle 24 and, similar to the previous minimum of solar cycle 23, it is expected that we will experience a sunspotless activity for around 2\,--\,3 years during the current solar cycle minimum. In order to access how the Earth's ionosphere would behave in the post-minimum of cycle 24, Janardhan et al. 2015 examined the correlation of the critical frequency in MHz of F-region (\emph{foF2}) of Earth's ionosphere with the smoothed sunspot number (SSN) for the time period 1994\,--\,2014, and reported a strong correlation with a correlation coefficient,  r of 0.96. From the estimation of the quantity \emph{foF2}, we would know the value of the minimum reflection cut-off frequency of the Earth's ionosphere. The authors estimated that the \(\emph{foF2}^{2}\) for the period 2008\,--\,2009, covering the duration of the minimum of cycle 23 when there is no sunspots altogether, was \(10 (MHz)^{2}\) implying that the maximum reflection cut-off frequency would be $\le$ 3.5 MHz. Janardhan et al. (2015) therefore reported that the period after 2020, with a prolonged low levels of night-time maximum reflection cut-off frequency well below 10 MHz, would help to obtain systematic ground based VLF radio observations. It is important to note that quiet-Sun conditions post 2020 would provide an excellent opportunity to perform VLF radio astronomy. Therefore, we propose and discuss a VLF radio spectrum explorer which is located at Mingantu station.

The system composition and specification of the VLF radio explorer is briefly described in Section 2, while the design of the radio explorer is discussed in Section 3.  In Section 4, scientific observations and analyses related to the radio spectrum explorer are presented, followed by conclusion.

%% Authors can give a citation as 'Michel et al. 1992'.
%% You may also use \cite, \citep and \citet for citation, and use Table~1 or Figure~1
%% and so forth. Using \ref and \label for cross-references of Tables/Figures
%% is a good way in adjusting/adding/removing text, tables or figures.

\section{System Composition and Specification}
\label{sect:syspec}

\begin{figure}[t]
\centering
\includegraphics[width=5.2in]{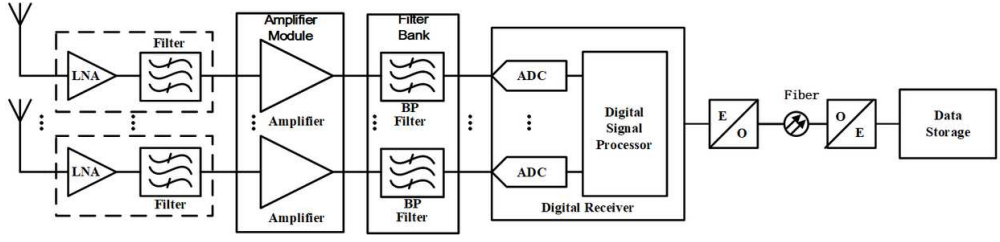}%
\caption{System schematic of the VLF radio spectrum explorer.
\label{fig:system}}
\end{figure}

The proposed VLF radio spectrum explorer is composed of four low frequency active antennas, two analog receiving modules, a GPS receiver, and a digital receiver combined with a storage array. Each antenna receives radio signals with two linear polarizations. The eight-channel radio frequency (RF) signals from the four antennas are amplified and filtered by an amplifier module and a filter bank separately. Then the radio signals will be digitized by Analog to Digital Converter (ADC) with suitable power levels and frequency bandwidths. In the digital receiver synchronized by a GPS receiver the sampled signals will be processed and stored for scientific studies. A systematical schematic of the radio explorer is shown in Figure \ref{fig:system}.

The upper limit of operating frequency of the instrument is purposefully set to 70 MHz. This would provide a common frequency window of 30--70 MHz with other available low frequency radio telescopes so as to perform joint observations. The joint observations would, in particular, be used for instrumental calibration purposes and common science case studies. The setting of upper limit frequency to 70 MHz would allow our radio observations free from the FM band (87--108 MHz) signals. Thus, we could avoid the strong RFIs, and it also allows the band-pass filter to have a high stop-band rejection for the sampling bandwidth of 80 MHz. The lower frequency limit is designed to be 1 MHz so as to prevent the strong MW and LW RFIs. In the whole frequency band, 2048 frequency channels are implemented to get the high resolution dynamic spectrum with the minimum integration time of 10 milliseconds. The system dynamic range of the instrument is about 35 dB. The main specifications are listed in the Table \ref{tab:spec}.
\begin{table}[b]
\begin{center}
\caption[]{Basic specifications of the VLF radio spectrum explorer.}
\label{tab:spec}
%%Please Capitalize the First Letter of Each Notional Word in table's caption
\begin{tabular}{ll}
\hline\noalign{\smallskip}
Antenna  & 5 m, \(4 \times 2 \) dipoles \\
Polarization & Dual linear \\
RF channel & 8 \\
Frequency range & 1\(\sim\) 70 MHz \\
Frequency channel & 2048 \\
Frequency resolution & \(\sim\) 39 kHz  \\
ADC & 12 bits @ 160 MHz \\
Time resolution & \(\geq\) 10 ms \\
Sky noise limited & 5 dB, \(8 \sim\) 65 MHz \\
System dynamic range & \(\geq 35\ \rm dB\)  \\
Sensitivity & \(\sim 10^{4}\) Jy @ 30MHz (10kHz,100ms) \\
\noalign{\smallskip}\hline
\end{tabular}
\end{center}
\end{table}

%________________________________________ Table 2: Use_of_the routines

\section{System Design}
\label{sect:sysdesign}

The whole system of the radio spectrum explorer mainly includes two parts, outdoor units and indoor units. The outdoor units include the active antennas, the analog receiving modules, the digital receiver, and the GPS receiver, which are hosted by an electronics cabinet. The indoor units include the monitoring and control computers, and the storage array. The observation data output by the digital receiver will be transmitted to the inside receivers via an optical fiber.

\begin{figure}[t]
\centering
\includegraphics[width=3.2in]{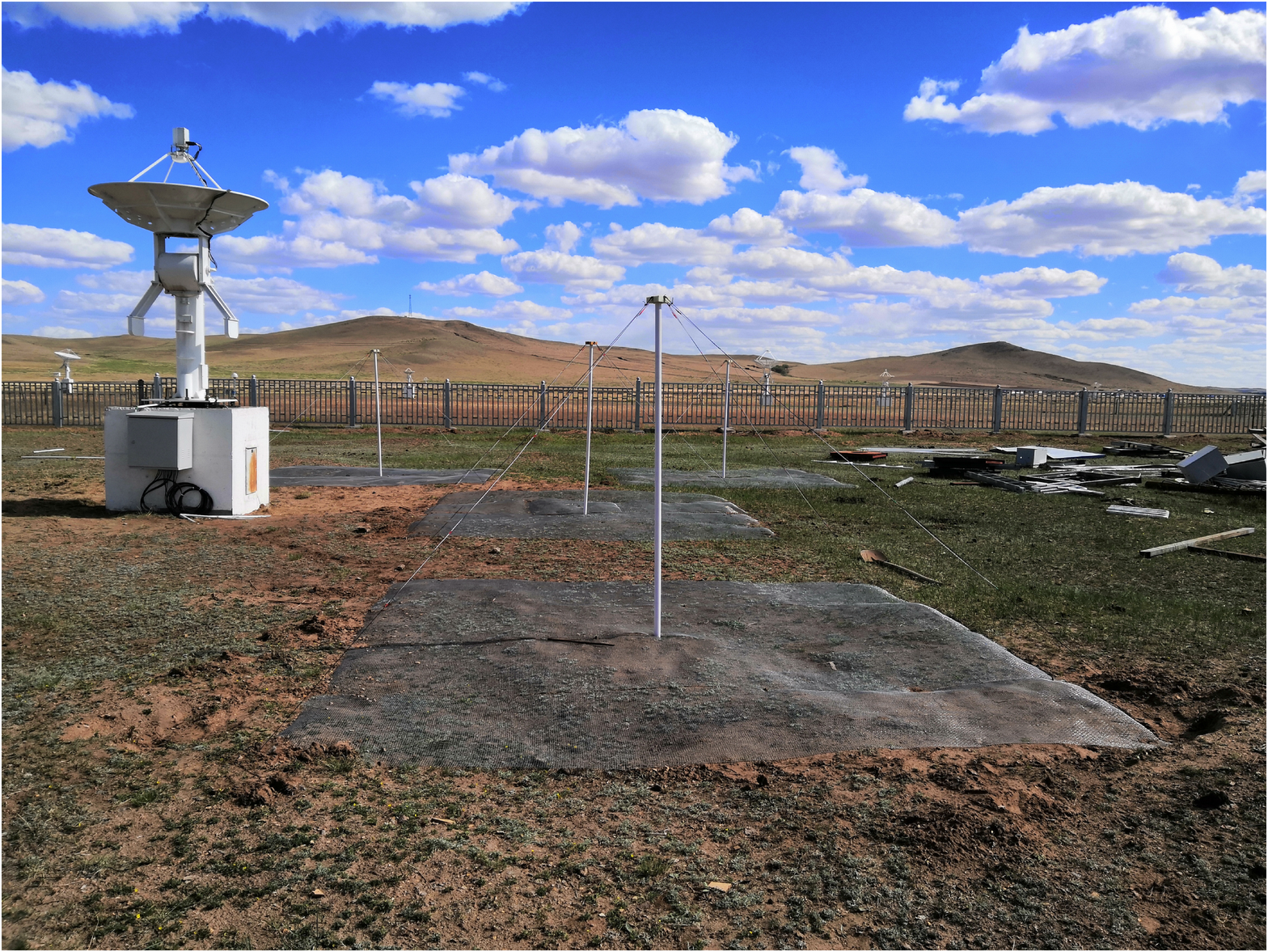}%
\caption{Radio spectrum explorer built at Mingantu Observation Station. It includes four inverted-V cross-dipole antennas standing above the ground metal meshes. The low noise amplifiers are at the top of the four supporting poles. For each antenna three cables inside the tubular pole are used to power the antenna and transmit its radio signals to the outdoor receiver, which is hosted by a RF-shielded cabinet hanging on the base block of the left parabolic antenna.
\label{fig:antenna}}
\end{figure}

\subsection{Antenna Configuration}
At very low frequencies setting up a resonant antenna is not feasible to receive the radio signal due to its long wavelength. In this scenario, we need an active antenna in order to obtain a good gain and a wide frequency band. The resonant frequency of the active antenna should be set  around 35 MHz so as to achieve a frequency range from 1 to 70 MHz. However, we set the resonant frequency of the antenna to be around 30 MHz. Thus, we could avail more observations at the low end of the frequency band to study in detail and explore the science at very low frequency regime. However, changing the resonant frequency to 30 MHz we make a trade-off between the sensitivity and the frequency band. Therefore, the length of the antenna is set to be around 5 meters. An inverted-V shaped dipole is designed as the receiving antenna. The antenna rests on a ground plane consisting of a metal mesh to avoid the reflected radio signals from the ground. The inverted V-shaped dipole antenna resting on a metal mesh is shown in Figure \ref{fig:antenna}.

\begin{figure}[t]
\centering
\includegraphics[width=\columnwidth]{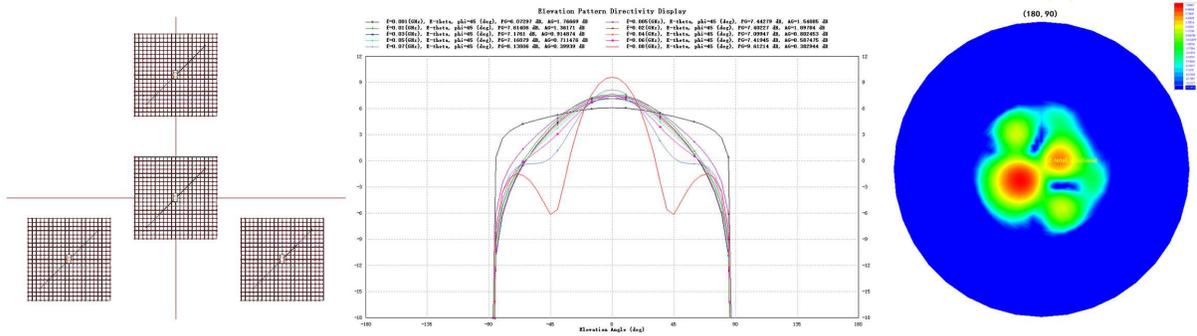}%
\caption{Antenna layout and simulations for the radio spectrum explorer. Left: Layout of the four inverted-V dipole antennas. Middle: The simulated radiation patterns of the single antenna. Right: The simulated beamforming pattern at 30 MHz for the four antennas.
 \label{fig:layout}}
\end{figure}

In order to have a wider field of view (FoV), the inverted-V dipole antenna is designed with an angle of \(109^{\circ}\) between the two monopoles. This, in principle, will broaden the radiation pattern of the antenna. Two perpendicular inverted-V dipoles are combined together to form a cross-dipole antenna as the receiving antenna element. The radio spectrum explorer consists of four inverted-V cross-dipole antennas. Three of them stand on the ground forming a regular triangle, the other one lie at the center of the triangle. All these three antennas have the same distance of 6 meters from the central antenna. The antenna layout is shown in Figure \ref{fig:layout}. The four antennas of the radio spectrum explorer form a mini array. So we can get different radiation patterns for the mini-array by the method of beam-forming.

As shown in Figure \ref{fig:layout}, the single inverted-V antenna has an E-Plane beam widths varying from 153.9 to 47.6 degree between 1 to 70 MHz. At lower frequencies, it can almost observe all the sky. Using detailed simulations, it is shown that the meshed ground plane improves the gain of the antenna at most of the frequencies. Although the distances between the four antennas are short-spaced, as shown in Figure \ref{fig:layout}, it is, however, still possible to obtain a rough pointing at higher frequencies via beam-forming.

In an active antenna, a low noise amplifier (LNA) connected with the antenna is used to receive and amplify the radio signals with low noise. In a 50-ohm system the LNA of the active inverted-V antenna can achieve a low noise figure of \(\sim 2.0\) dB at most of the operating frequency band. Also, the LNA allows a matching between the antenna and the transmission cable. In the design of this LNA, a two-stage amplifier (\cite{chen2018}) is employed to obtain a reasonable gain and low noise. While the use of the LNA reduces the noise contributions, the available gain of the active antenna is decreased as well due to mismatching between the active antenna and LNA. For one inverted-V cross-dipole antennas, two LNAs are used and placed on one printed circuit board (PCB) combined together. The PCB is kept inside a 3D printing box, which is fixed on the top of the antenna supported by a pole as shown in Figure \ref{fig:antenna}.

\begin{figure}[b]
\centering
\includegraphics[width=\columnwidth]{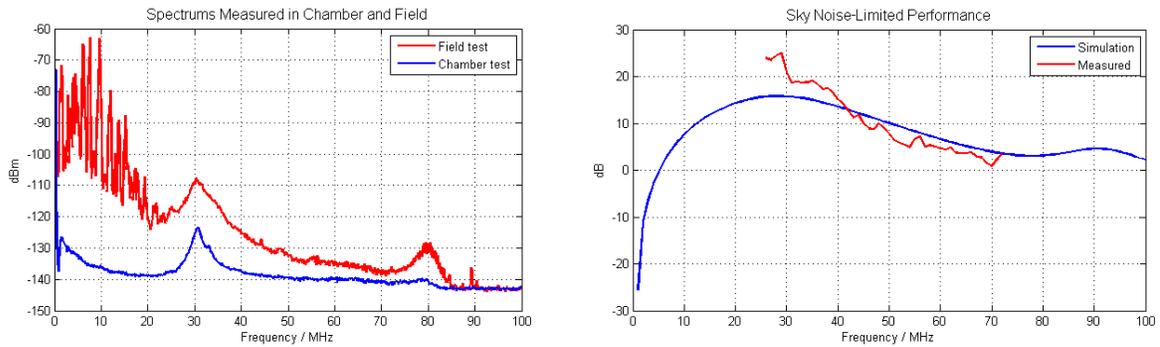}%
\caption{Sky noise-limited values as measured for an inverted-V dipole antenna. Left: Calibrated noise power density as measured in the chamber and at Mingantu observation station. Right: The performance of the simulated and measured results of the sky noise-limited values.
\label{fig:snlp}}
\end{figure}

At very low frequencies, the sky noise limited performance is a primary requirement for the antenna system to carry out observations with high spectral and temporal resolutions. This shows how large the sky noise is during the output of the antenna as compared to the noise from the antenna system itself. In order to meet this requirement, the LNA is designed to achieve a sky noise-limited performance within each operational frequency band. Using a sky temperature model (\cite{heino2009}), we simulated the sky noise limited performance of the active antenna. As shown in Figure \ref{fig:snlp}, it is seen that an approximate of 5 dB sky noise limitation is achieved between 8 and 65 MHz. Using a method proposed in (\cite{chen2018}), we calculated the sky noise-limited performance for the active inverted-V antenna inside the chamber and in the field as well.  From Figure \ref{fig:snlp} it is clear that the measured results match well with the simulation results. However, it is noticed that above 42 MHz the measured results differ from the simulation results. It is probably caused due to the difference in the noise level of LNA as estimated by the measurement and the simulation. Inspite of this, it is evident that the measured results meet the required antenna noise level needed for our operational frequency range and clearly also illustrate the performance of the inverted-V dipole antenna designed for the radio explorer. It is to be though noted that we did not measure the sky noise-limited performance below the frequency of 25 MHz due to the RFI impact.

\subsection{Analog Receiving Module}
As mentioned earlier, an amplifier module following the antenna is employed to amplify the radio signal in order to make its power suitable for sampling. Due to the low available gain of the antenna as a result of mismatching with the amplifier, we consider to reduce the receiver noise after the antenna to improve the system sensitivity. The amplifier has a gain of around 22 dB and a low noise level of 2 dB. This, in turn, significantly reduces the noise contribution of the digital receiver. In order to avoid the signal aliasing in ADC, we use an anti-aliasing bandpass filter to achieve the operating frequency band of \(1 -70\) MHz.

\begin{figure}[h]
\centering
\includegraphics[width=\columnwidth]{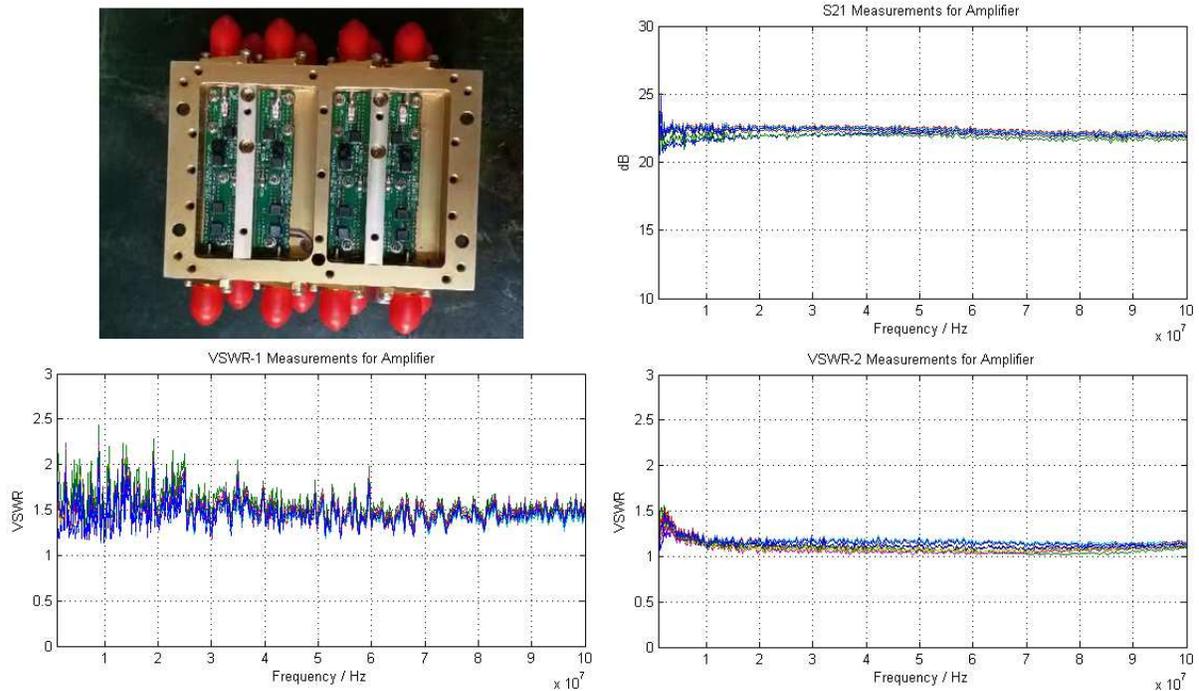}%
\caption{Amplifier measurements, different colors represent different amplifiers. Upper-Left: The amplifier module. Upper-right: The measured S-parameter S21, it shows the gain of the amplifier. Lower-left: The measured Voltage Standing Wave Ratio (VSWR) at the input port. Lower-left: The measured VSWR at output port.
\label{fig:amp}}
\end{figure}

\begin{figure}[h]
\centering
\includegraphics[width=\columnwidth]{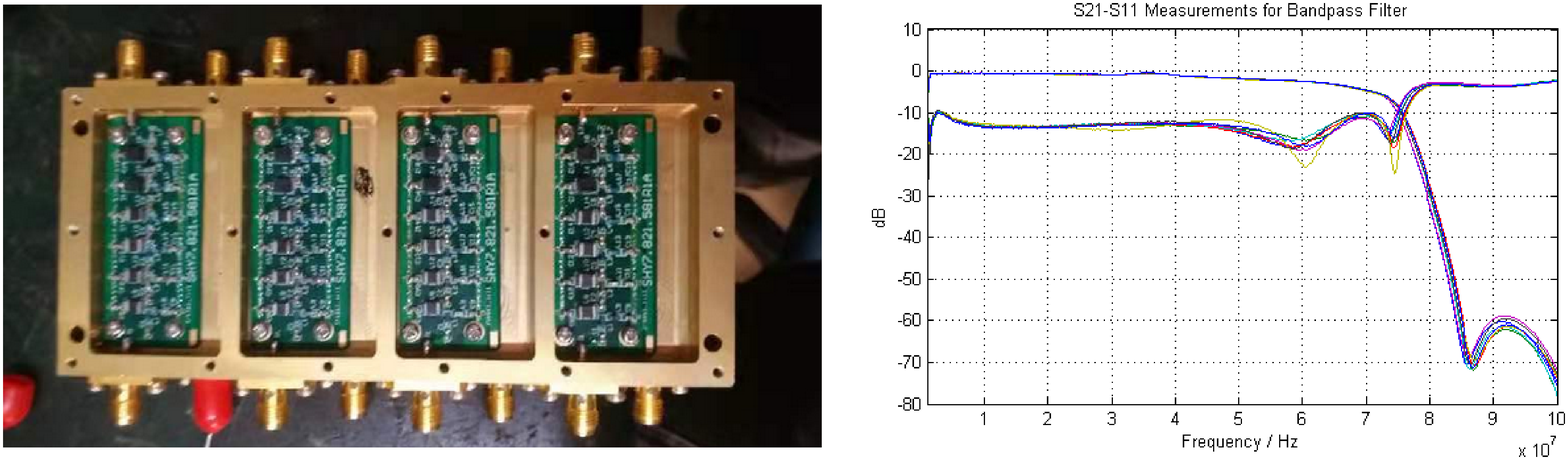}%
\caption{Bandpass filter measurements, different colors represent different filters. Left: The filter bank. Right: The measured S-parameter S21 showing the the frequency response of the filter, and S11 showing the reflection coefficient at the input port. Since the symmetrical characteristic of the filter, only half of parameters are plotted.
\label{fig:filter}}
\end{figure}

As shown in Figure \ref{fig:amp}, the gain of the amplifiers is about 22 dB, and the VSWR are less than 2 and 1.5, respectively, at input and output ports. We can also see that the measurement results coincide with each other among the amplifiers at most of the frequency range. Figure \ref{fig:filter} plots the measurement results for the filter bank. As we can see that the 3 dB bandwidths of the filters are from 1 to 68 MHz, it is slightly narrower than the design. The measurements show the filters accord with each other well. From both Figures \ref{fig:amp} and \ref{fig:filter}, it can be seen that there is lesser difference for the same measurements of both the amplifier and filter. This is mainly attributed to the non-identical characteristic of the electronic components.

\subsection{Digital Receiver}
For a low frequency radio array, the digital receiver is a core instrument. The digital receiver mainly perform all the required signal processing in the digital domain. In our proposed VLF radio spectrum explorer, the digital receiver used would sample the received signals by the active antenna with a 12-bit ADCs operating at 160 Million Samples per Second (MSPS). We then use a 4096-point Fast Fourier Transform (FFT) to channelize the received signals from all the four antennas. In this way,  we can obtain 2048 sub-band signals with a bandwidth of \(\sim39\) kHz. Each sub-band signal would be integrated next for at least 10 milliseconds, which can be changed as per requirement. Meanwhile, a piece of 4096-point raw sampled data would be sent for output in a single integration time period for all the antennas that can further be used for more flexible offline processing. For all the four dual-polarized dipole antennas, the maximum output data rate obtained is around 12.5 MB/s.

\begin{figure}[b]
\centering
\includegraphics[width=\columnwidth]{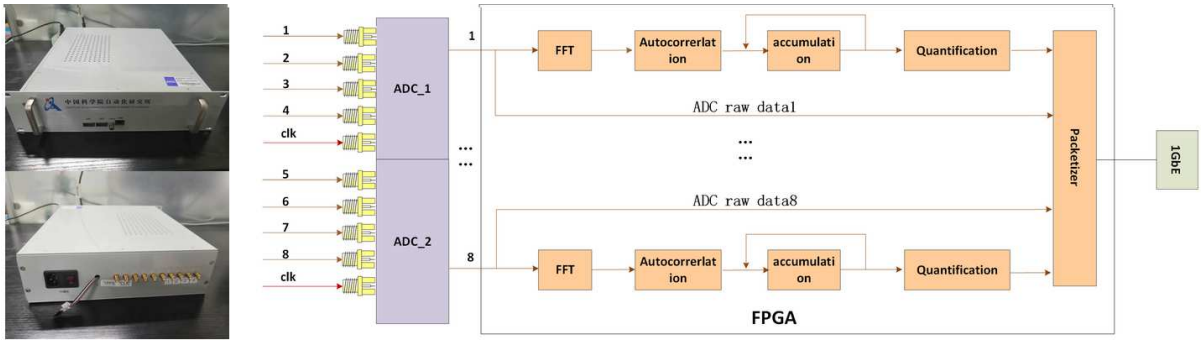}%
\caption{Digital receiver and its schematic diagram. The signal processing is implemented in two different methods to provide the spectrum and raw data.
\label{fig:digtal}}
\end{figure}

\subsection{System Control, Monitoring and Storage}

\begin{figure}[h]
\centering
\includegraphics[width=3.5in]{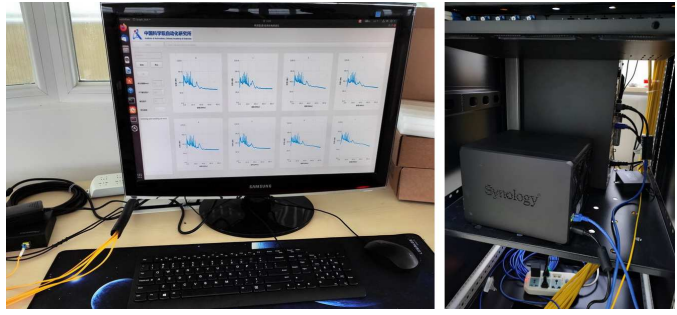}%
\caption{System control, monitoring and storage for the radio spectrum explorer. Left: the control and monitoring interface displaying on a screen locating in an EM shielding observation room. Right: the control computer and disk array in a high-level EM shielding room. The screen and computer are connected with each other via a fiber to avoid the RFI leaking.
\label{fig:control}}
\end{figure}

In the radio spectrum explorer, the observation data produced by the digital receiver would be transmitted to an indoor receiver via optical fiber. As shown in Figure \ref{fig:control}, a high-performance computer located in a high-level EM shielding room is used to receive the data and store them in a disk array. Meanwhile, the computer also control the digital receiver and monitor the observation results in a real time. For the convenience of observing, the system control and monitoring can be achieved remotely on any other authorized computer with the internet connection.

\subsection{System Analysis}
In order to analyze noise distribution for the radio spectrum explorer, we define the antenna noise and receiver noise with a reference plane between the antenna and the LNA used in (\cite{chen2019}). The antenna noise includes the background sky noise and the thermal noise from the antenna, while the receiver noise includes the noise from the LNA, analog and digital receiving units. We use the temperature model in (\cite{heino2009}) and simulate the different noise contributions from the whole system. The noise figure of the amplifier after the antenna is around 1.0 dB. Thus, its noise temperature is around 75 K. As the gain of the amplifier is 22 dB and the available gain of the LNA is about 20 dB, we can ignore the equivalent noise of the amplifier and digital receiver at the reference plane.

\begin{figure}[h]
\centering
\includegraphics[width=\columnwidth]{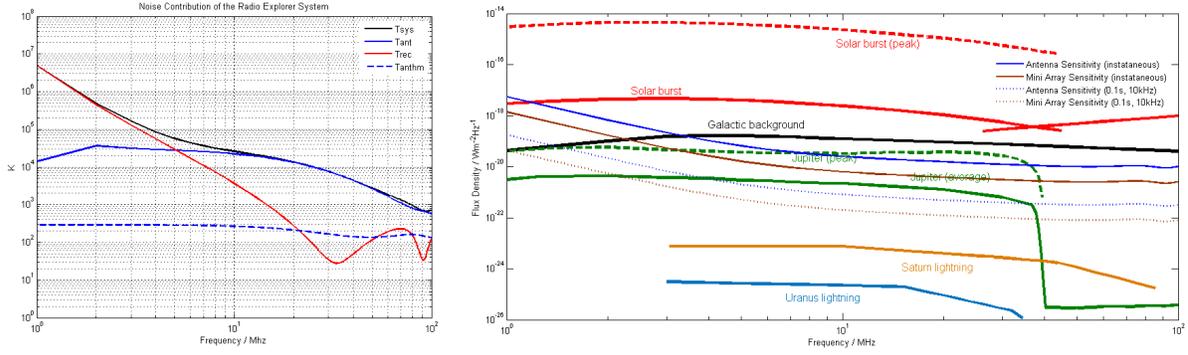}%
\caption{Left: Simulated noise contributions of the receiving chain for the radio spectrum explorer. The black, blue, red, and dashed blue plots represent respectively the system noise, the antenna noise, the receiver noise and the antenna thermal noise. Right:Flux densities of radio emissions from different cosmic objects along with the antenna and array sensitivities are plotted. For plotting the different cosmic radio emissions, we used the data in (\cite{zarka2012}), while for the sensitivities of the antenna and array system, we performed simulations with different integration time and bandwidth.
\label{fig:noiseflux}}
\end{figure}

The simulated noise contributions of the system at the reference plane are plotted in Figure \ref{fig:noiseflux}. It can be seen that the antenna noise exceeds the receiver noise above 7 MHz. Generally, the sky noise is very high and the antenna thermal noise is too small. Hence, it can be concluded that the sky noise is always larger than the receiver noise. Further analysis shows that the sky noise is at lease 5 dB more than the receiver noise above 8 MHz, which agrees well with the sky noise-limited result shown in Figure \ref{fig:snlp}. Similar to the simulation of noise contribution, the sensitivity of the antenna as well as the mini array was also simulated and plotted in Figure \ref{fig:noiseflux}. From Figure \ref{fig:noiseflux}, it is clear that the antenna and mini array sensitivities lie well below the peak and average radio emission fluxes from the Jupiter. We thus conclude that even the single antenna sensitivity is good enough to detect the radio emission from the Jupiter. In the other hand, the solar radio bursts, which generally produce strong radio emission at very low frequency range, could be as well detected by the antenna and mini array.

The system dynamic range for the radio spectrum explorer depends on all the active devices such as the active antenna, amplifier and digital receiver. Simulations of the system show that, in absence RFI, the total power of the radio signal received by the antenna is about -48 dBm. This roughly agrees with the measurement carried out at the instrument site. Similarly, the gain of the amplifier is 22 dB and its input power of 1 dB compression point is around -8 dBm. Thus, we estimate that the net signal power at the ADC input would be -26 dBm. If we use a 12-bit ADC with a full-scale of 10 dBm, then the net signal power at the ADC input would fill only 6 bits.  The rest 6 bits could be used for obtaining the high dynamic range signals. Thus, in absence of the RFIs, the radio spectrum explorer could have a system dynamic range of around 36 dB.

\section{Scientific Observations}
\label{sect:sciences}
The system design as aforementioned for the radio spectrum explorer would make it a competent instrument at VLF, which can be useful for carrying out studies of various cosmic objects and phenomena such as solar radio bursts, planetary radio emissions and joint low frequency observations with other available facilities.

\subsection{Expected Scientific Studies}
A round clock monitoring of solar activity is required to understand space weather in near-Earth space. In particular, changes pertaining to eruptions of solar flares and coronal mass ejections (CME) are crucial as such phenomena are not well studied in near-Earth space. One way to quantify the changes is the study of radio burst activity associated with them. There are several types of burst activity, but the two burst Types are of particular interest: Type II and III. Type II bursts are produced by fast driving shocks associated with the flares or CMEs, while Type III bursts are a result of propagating electron beams in solar corona and interplanetary space. These bursts are produced near the local electron plasma frequency or/and its harmonics and as the plasma frequency depends on the electron number density, the source position of each burst corresponds to a certain height in the corona. Using our proposed VLF radio spectrum explorer, which can be used as a highly effective solar monitoring system, we target to identify these Type II and III burst types from the solar dynamic spectra. In coordination with LOFAR and LWA solar imaging spectroscopy, we will study the plasma processes associated with these bursts to prepare a density diagnostics of the energetic electrons at different heights of the corona. This, in turn, will enable the study of density diagnostics of the different structure of CMEs as well as determining their velocity and acceleration. Also, through observation of absorption bursts from solar dynamic spectra of VLF radio spectrum explorer, we plan to locate the source region of CMEs and plasma characteristics associated with the emerging magnetic flux channels. Hence, this instrument can effectively be used to study the propagation and evolution of CMEs as well as the non-thermal electron beams associated with the flares in the interplanetary space through monitoring of solar radio burst activity.

\begin{figure}[t]
\centering
\includegraphics[width=\columnwidth]{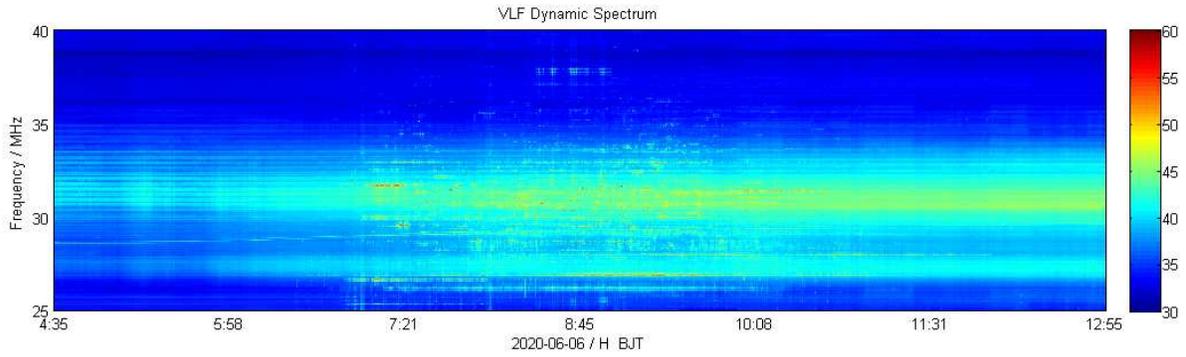}%
\caption{Spectrum of ionospheric low frequency radio emission observed by the radio spectrum explorer on June 6, 2020.
\label{fig:solarburst}}
\end{figure}

In addition to solar activity, the VLF radio instrument can also be used to study planetary radio emissions. It is known that the planet Jupiter emits strong radio emissions ($\sim$MJy), which are point-like, fully polarized and quasi-permanent in 0.3--3 MHz range. In particular, the Jupiter emissions can be well measured in the decameter (30 MHz) range (\cite{zarka2004},~\cite{nigl2007}). Thus, the Jupiter emissions can be useful for ground calibration for the radio spectrum explorer. While the planetary radio emissions from Saturn and Uranus can be used to study quasi-continuous radio emissions and their seasonal effects.

\begin{figure}[b]
\centering
\includegraphics[width=\columnwidth]{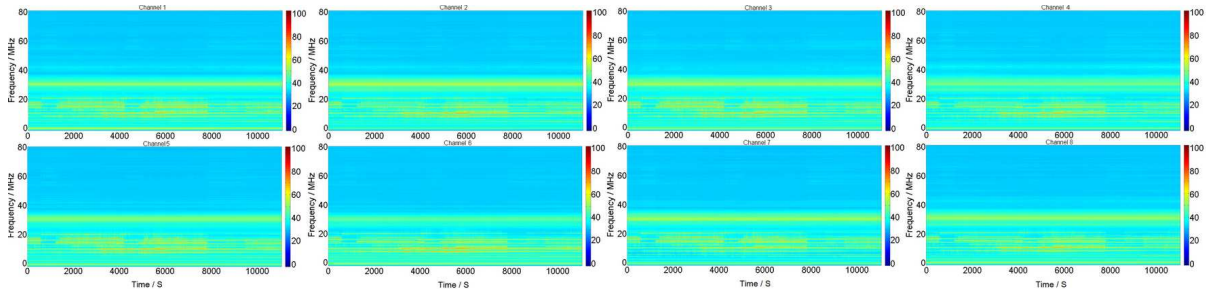}%
\caption{Dynamic spectrum observed by the four dual-polarization antennas of the built radio spectrum explorer on May 31, 2020.
\label{fig:8chn}}
\end{figure}

The ground VLF radio spectrum explorer can be used to perform joint observations with the Netherlands-China low frequency explorer (NCLE), which is a radio instrument onboard the Chinese Chang'E 4 lunar mission (\cite{boonstra2017}) to perform radio observations in the frequency range 0.08--80 MHz. The key science objectives of the NCLE, that operates in the similar frequency range like our VLF radio spectrum explorer, are overlapped with that of our instrument, such as measuring radio emission from different planets in our solar system, studying Earth's ionospheric fluctuations, and solar radio activity and space weather. We, therefore, target for the joint spectrum observations from the VLF radio spectrum explorer and the NCLE instrument in the frequency range 1-70 MHz for the studies of planetary radio emission, solar radio bursts, and space weather in near-Earth space.

\subsection{Observing Modes}

\begin{figure}[t]
\centering
\includegraphics[width=\columnwidth]{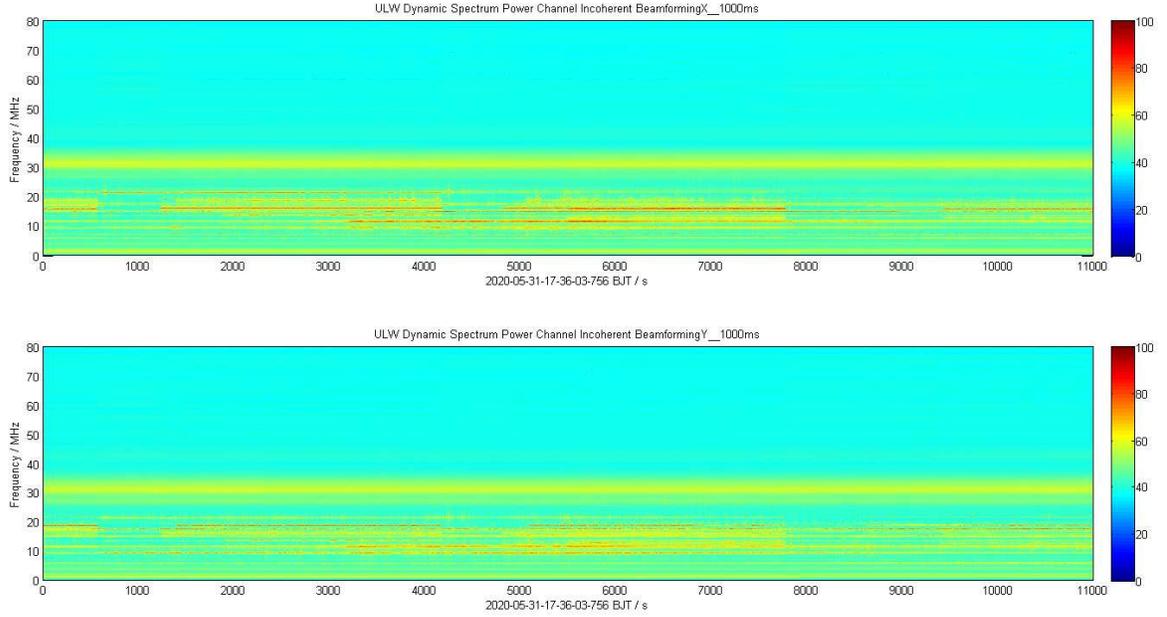}
\caption{Dynamic spectrum observed with the incoherent beam pointing at the zenith on May 31, 2020.
\label{fig:beamforming}}
\end{figure}

As mentioned in Section \ref{sect:sysdesign}, this radio spectrum explorer contains four VLF radio antennas, and the digital receiver will produce two different data outputs including the spectrum data and raw sampling data, which make it capable to do the observations in different working modes. If the four antennas work independently, each of them can be considered as a spectrometer as shown in Figure \ref{fig:8chn}. These four antennas can also work together as a mini array to improve the sensitivity. For the spectrum data, we can sum the same polarized signals of the four antennas together to get the combined signal with an incoherent beam, as shown in Figure \ref{fig:beamforming}. Suppose the four antennas are identical, the incoherent beam is the same with the beam of one single antenna. In this way, the sensitivity of this radio spectrum explorer will improve by the times of square-root of the antenna number. For the raw sampling data, we can do the coherent beam-forming by summing the four-antenna signals together with phase shifting. Using this way, this radio spectrum explorer can observe the Sun or other radio sources with rough tracking, though the formed beam is big at low frequencies. Besides, the raw data can also be used to make the cross correlations between different antennas for signal chain calibrations or locating radio source positions, etc.

\section{Conclusions}
\label{sect:conclusion}
In the present work, the design for a radio spectrum explorer built in Mingantu station in China has been discussed. It operates in the VLF range of 1--70 MHz, and will carry out the VLF radio observations of Sun, different planets as well as different other celestial objects. This instrument is currently installed at the Mingantu site, and using in house simulations and direct observations of the radio sky measurements at the site. It is shown that the instrument is a sky noise-limited system at most of the operating frequency band taking into account all the possible noise contributions. The instrument has a very good sensitivity and can provide a reasonable dynamic range. Currently,the VLF radio spectrum explorer is in operation and has been used to produce VLF radio dynamic spectra. The radio spectra obtained can be used to perform several scientific studies, including the study of solar activity and space weather through a continuous monitoring solar radio burst activity associated with the energetic electrons from the solar flares and CMEs, planetary radio emissions, and carrying out joint observations as well with the VLF payloads onboard Chang'E-4 mission.

Using the instrument, we, particularly, can obtain radio dynamic spectra. However, in order to obtain the
complete information about the plasma processes associated with the radio emission at this VLF frequency range, we need to have interferometric observations as well to produce radio images. In this direction, in future, we are planning for another larger VLF radio array consisting of 36 antennas within 3 km to be built soon at the same site. The upcoming interferometer would be able to image solar radio bursts and can be used to produce the low frequency radio sky map with a reasonable spatial resolution. Using the imaging spectroscopy, we can reveal useful information about the source regions of the radio emission, and their distribution based on the density gradient.

\section*{Acknowledgements}
This work was funded by the National Natural Science Foundation of China(NSFC) 11573043, 11790305 and 11433006, National Key R\&D 278 Program of China (2018YFA0404602), the CE-4 mission of the Chinese Lunar Exploration Program: the Netherlands-China Low Frequency Explorer (NCLE), and Chinese Academy of Sciences (CAS) Strategic Priority Research Program XDA15020200 are acknowledged. The authors would like to thank Mr. Jianxi Ren and Ms. Zhijun Chen from NAOC for the help in building this radio explorer.

\bibliographystyle{spbasic}      % basic style, author-year citations
%\bibliographystyle{spmpsci}     % mathematics and physical sciences
%\bibliographystyle{spphys}{ieeetr}       % APS-like style for physics

\begin{thebibliography}{39}

\bibitem{alexander1975} Alexander, J. K., et al., 1975, AAP, Vol. 40, pp. 365

\bibitem{bergman2009} Bergman, J. E. S., Blott, R. J., Forbes, A. B., Humphreys, D. A., Robinson, D. W., Stavrinidis, C., CEAS 2009 ¨C European Air and Space Conference. 26-29 October, 2009, Manchester, U.K.

\bibitem{boonstra2017} Boonstra, A. J., et al., Porceeding of 32nd URSI GASS. Montreal, 19-26 August 2017.

\bibitem{boonstra2016} Boonstra, A. -J., et al., 2016 IEEE Aerospace Conference. Yellowstone Conference Center, Big Sky, Montana, Mar 5-12, 2016.

\bibitem{chen2019} Chen, L., Yan, Y.(2019)., Radio Science, 2019, 54.

\bibitem{chen2018} Chen, L., Aminaei, A., Gurvits, L. I., Wolt, M. K., Pourshaghaghi, H. Reza., Yan, Y., Falcke, H., Experimental Astronomy, 2018, Vol. 45, Issue 2, pp.231-253.

\bibitem{ellingson2009} Ellingson, S. W., Clarke, T. E., Cohen, A., Craig, J., Kassim, N. E., Pihlstrom, Y., Rickard, L. J., Taylor, G. B., Proc. IEEE, 2009 Vol. 97, No. 8, 1421

\bibitem{gopalswamy2014} Gopalswamy, N.; Makela, P.; Yashiro, S., 2014 United States National Committee of URSI National, 2014 vol., no., pp.1-1, 8-11 Jan.

\bibitem{haarlem2013} Haarlem, M. P. V., et al. (2013)., Astronomy \& Astrophysics, 2013, Vol. 556, No. 2.

\bibitem{herman1973} Herman, J. R., Caruso, J. A., Planet. Space Sci., 1973, Vol. 21, 443-461.

\bibitem{janardhan2015} Janardhan, P., Bisoi, S.K., Ananthakrishnan, S., Sridharan, R., Jose, L., Sun and Geosphere, 2015, vol.10, no.2, p.147

\bibitem{heino2009} Jester, S., Fackle, H., New Astronomy Review, 2009, 53, 1

\bibitem{ji2017} Ji Y.-C, Zhao B.,Fang G.-Y, et al., Journal of Deep Space Exploration, 2017, 4 (2): 150

\bibitem{kaiser1996} Kaiser, M. L., Desch, M. D., Bougeret, J. L., Manning, R., Meetre, C. A., Geophysical Research Letters, 1996, Vol. 23, Issue 10, 1287

\bibitem{konovalenko2016} Konovalenko, A., Sodin, L., Zakharenko, V. et al., Exp Astron, 2016, 42: 11.

\bibitem{nigl2007} Nigl A., Zarka P., Kuijpers J., Falcke H., Bahren L., Denis L.,
    A\&A, 2007, 471 (2007), pp. 1099

\bibitem{zarka2012} Zarka, P., Bougeret, JL, Briand, C., Cecconi, B., Falcke, H., Girard, J., Griebmeier, JM, Hess, S., Wolt, M.,Konovalenko, A., Lamy, L., Mimoun, D., Aminaei, A., Planetary and Space Science, 2012, 74. 156

\bibitem{zarka2004} Zarka, P., Planetary and Space Science, 2004, 52( 15), 1455

\end{thebibliography}


\label{lastpage}

\end{document}